# Superconducting proximity effect in inverted InAs/GaSb quantum well structures with Ta electrodes

Wenlong Yu,[†] Yuxuan Jiang,[†] Chao Huan,[†] Xunchi Chen,[†] Zhigang Jiang,*,[†] Samuel D. Hawkins,[‡] John F. Klem,[‡] and Wei Pan*,[‡]

[†]School of Physics, Georgia Institute of Technology, Atlanta, GA 30332, U.S.A.

[‡]Sandia National Laboratories, Albuquerque, NM 87185, U.S.A.

ABSTRACT

We present our recent electronic transport results in top-gated InAs/GaSb quantum well hybrid structures with superconducting Ta electrodes. We show that the transport across the InAs−Ta junction depends largely on the interfacial transparency, exhibiting distinct zero-bias behavior. For a relatively resistive interface a broad conductance peak is observed at zero bias. When a transparent InAs−Ta interface is achieved, a zero-bias conductance dip appears with two coherent-peak-like features forming at bias voltages corresponding to the superconducting gap of Ta. The conductance spectra of the transparent InAs−Ta junction at different gate voltages can be fit well using the standard Blonder-Tinkham-Klapwijk theory.

Although the physics of a semiconductor (Sm) heterostructure in contact with a superconductor (S) has been studied extensively for several decades, the subject has seen renewed interest with the possibility of realizing Majorana fermions in strong spin-orbit coupled Sm−S hybrid structures[1-7]. The emergence of Majorana fermions in solid-state systems may have profound implications in quantum computation[8-10] and it has triggered an avalanche of research activities[1-7, 11-14]. Recently, two-dimensional topological insulators (or quantum spin Hall insulators) have attracted a great deal of attention due to being a promising material system for realizing Majorana fermions[9]. Among the probable quantum spin Hall insulators, the type-II InAs/GaSb quantum well (QW) structure[15] is a rising candidate. It was shown that with an inverted band structure this material can support dissipationless time-reversal symmetry protected spin edge channels and display the quantum spin Hall effect[16, 17]. When in proximity to a superconductor, it can host Majorana zero-modes[1, 18, 19], providing a unique platform for investigation. Furthermore, InAs QWs are particularly suited for fabricating hybrid Sm−S devices, owing to the low electron effective mass and high mobility as well as the low Schottky barrier interface to superconductors[20]. Therefore, it is of fundamental interest to study the proximity-coupled InAs/GaSb hybrid structures.

In this Letter, we report on a systematic study of the proximity effect in top-gated InAs/GaSb QWs in contact with a superconducting Ta electrode. We find that the electronic transport across the InAs−Ta junction exhibits distinct zero-bias behavior, either a conductance ($dI/dV$) peak or dip, depending on the interfacial transparency. We show that although a zero-bias $dI/dV$ peak is often seen across relatively resistive junctions[21], the conductance spectra of transparent interface recover the characteristic lineshape of Blonder-Tinkham-Klapwijk (BTK) theory[22], with a $dI/dV$ dip occurring at zero bias and two coherent-peak-like features at bias voltages corresponding to the superconducting gap of Ta.

The InAs/GaSb QW bilayer studied in this work was grown by molecular beam epitaxy, following the recipe described in ref [15]. The inset of Figure 1a illustrates the schematic of the QW structure, in which a 150 Å InAs QW and a 40 Å GaSb QW are sandwiched between AlSb barriers. Two-dimensional electrons and holes are expected to coexist in this system and to be confined in the InAs and GaSb QWs, respectively. Transport characterization of the material has been reported previously[23]. The integer quantum Hall effect is found fully developed in both the electron- and hole-dominated regimes, attesting to the high quality of the material. To fabricate Sm−S hybrid devices, we first define a 20 μm x 24 μm InAs/GaSb mesa using conventional photolithography techniques and wet chemical etching. Ammonium hydroxide is used to etch AlSb and GaSb selectively, while citric acid/hydrogen peroxide solution is used to remove InAs. Normal-metal Au/Ti (200/10 nm thick) electrodes are then patterned to connect with both the InAs and GaSb QWs at the four corners of the mesa, and superconducting Ta (70 nm thick) electrodes are sputter-deposited directly on top of the InAs layer (forming InAs−Ta junctions) after removing the InAs cap layer, the 500 Å AlSb top barrier and the GaSb QW. The transition temperature $T_c$ of the Ta electrodes is ~1.4 K. Finally, a 100nm-thick $SiO_2$ dielectric layer is deposited using plasma-enhanced chemical vapor deposition and an Au/Ti top gate is fabricated on $SiO_2$ and covering the entire InAs/GaSb mesa. The inset of Figure 1c shows a scanning electron microscope image of one of the three devices we have studied. The minimum distance between two Ta electrodes is ~2 μm.

From now on, we focus on the transport across Sm−S junctions with respect to external parameters such as carrier density (by tuning the top-gate voltage $V_{tg}$), temperature, and magnetic field. The differential resistance d$V$/d$I$ (or conductance d$I$/d$V$) of the junction is measured either using a dc method (*i.e.*, $I-V$ measurement) and then calculated numerically or using a standard dc + ac lock-in technique with a typical ac current of 50 nA. Consistent results are obtained between these two methods. Three-terminal configuration is used to measure Sm−S junctions. For example, the top InAs−Ta junction in the inset of Figure 1c can be measured by sending a current from $I^+$: 2 to $I^-$: 6 while recording the voltage between $V^+$: 1 and $V^-$: 5. Since the separation between the two superconducting electrodes in our devices is always much greater than the correlation length of Ta (only a few tens of nm)[24], no supercurrent is observed across the Ta−InAs−Ta junction and no cross-talk effect is expected between the transport across the two InAs−Ta junctions.

Figure 1a shows the d$V$/d$I$ spectra of two InAs−Ta junctions as a function of the dc current $I_{dc}$. Each spectrum is normalized to its high-bias value $R_N$ = (d$V$/d$I$)$_N$ at $I_{dc}$ = 100 $\mu$A, which consists of the normal-state resistance of the Ta electrode ($R_{N,Ta}$) and that of the junction. The sharp d$V$/d$I$ peaks at high bias are likely due to the current-driven destruction of superconductivity in the Ta electrode[25, 26] and its position indicates the corresponding critical current. We find that the d$V$/d$I$ peaks become weakened and move towards zero bias with increasing temperature, and eventually vanished around $T_c$ (data not shown). The position of the peaks depends on the interfacial transparency and the applied $V_{tg}$. Larger critical current is observed for a more transparent InAs−Ta junction (red curve in Figure 1a) and for positive $V_{tg}$ voltages corresponding to higher electron concentration in InAs/GaSb QWs. A great advantage of InAs/GaSb-based Sm−S hybrid devices is that the charge carriers are continuously tunable from purely electron to two-carrier hole dominated regime through the charge neutrality point[17, 23, 27-29], which is at $V_{tg}$ ≈ -4.8 V for our devices. Such a tunability was not achieved in earlier studies of InAs−Nb junctions[21], in which case modulation doping with Te was used to vary the electron concentration in different samples.

The charge transport across Sm−S junctions is often described by the interfacial conductance spectra, i.e., $dI/dV$ versus dc bias $V_{bias}$. For the InAs−Ta junction with a zero-bias resistance ~38 Ω at $V_{tg} = 0$ V (inset to Figure 1b), the $dI/dV$ spectra exhibit a pronounced peak at zero bias as well as two satellite dips below the normal-state conductance at high-bias voltages, consistent with that reported in previous works[21]. We note that an anomalous zero-bias conductance peak is ubiquitous in transport measurement of superconducting hybrid structures but its origin is often debated. For instance, Sheet *et al.*[30] argue that heating effects may cause a spurious conductance peak at zero bias, as the increase in bias voltage may lead the local current to exceed the critical current, giving rise to a $V_{bias}$-dependent decrease in $dI/dV$ (due to the resistance increase at the transition). However, if this mechanism is dominant in our measurements, one would expect the $dI/dV$ to exhibit a weak magnetic-field dependence at zero bias, when the heating effects vanish and the Ta electrode remains in the superconducting state. In this scenario, the effect of magnetic field is just to reduce the critical current, thus narrowing the width of the zero-bias conductance peak with increasing magnetic field. These expectations are in clear contrast with the magnetic-field dependent measurements shown in Figure 3(b) and 3(d) for a relatively resistive junction, where in the low magnetic field regime the differential conductance decreases substantially with increasing field. Moreover, via examining the $V_{tg}$-dependence of the zero-bias conductance (inset to Figure 1b), one can further argue that the transport across the junction is not in the thermal regime, where the $dV/dI$ is expected to be dominated by the Maxwell resistance proportional to the bulk resistivity of the QWs[30]. For our devices, when the InAs/GaSb QWs are tuned close to the charge neutrality point ($V_{tg} = -4.8$ V), the system would become more resistive than the case of $V_{tg} = 0$ V or 3.6 V, leading to a smaller value of zero-bias conductance (larger Maxwell resistance). This is again in contrast with our experimental data.

A better interpretation of our observations for the relatively resistive junctions is the original proposal of ref [21], inspired by the work of van Wees *et al.*[31]. Following their interpretation, the anomalous zero-bias conductance peak is attributed to a multiple reflection process between the InAs−Ta interface and the AlSb back wall of the InAs QW, while the unusual broadening of the peak is due to the presence of a spatially nonuniform local bias between the InAs layer and the Ta electrode. The microscopic process underlying the proximity effect at the InAs−Ta interface is the so-called Andreev reflection[32]: at temperatures well below the superconducting gap of Ta, an electron in the InAs QW cannot transmit through the InAs−Ta interface, but rather is reflected as a coherent hole with simultaneous generation of a Cooper pair in Ta. Due to the long mean free path of charge carriers in our devices and the thinness of QWs, the electrons and Andreev holes may undergo multiple normal reflections between the two walls of InAs QW, giving rise to "excess" conductance at zero bias. This process is phase-coherent, and therefore sensitive to magnetic field and back-scattering. In this scenario, when the system is tuned towards charge neutrality, the scatterers in the InAs QW would become less screened and one would expect a substantial increase in excess conductance, which is consistent with our measurements shown in the inset of Figure 1b.

Another prediction from the theory of van Wees *et al.*[31] is the vanishing excess conductance for an ideal Sm−S interface with the assumption of classical scattering in the Sm. Below we will demonstrate that this behavior is indeed evidenced in the d$I$/d$V$ spectra of more *transparent* InAs−Ta junctions and BTK-like spectral lineshape is recovered with the extracted superconducting gap of Ta consistent with that from temperature dependent transport measurements. Here, the highly transparent Sm−S junctions are achieved by choosing Ta as the superconducting material and more importantly by minimizing the interval between the ammonium hydroxide etching and Ta sputter deposition, i.e., minimizing the exposure time of fresh InAs surface to air. In Figure 2a we plot the d$I$/d$V$ spectra of a transparent InAs−Ta junction in a relatively large dc bias voltage range and at different $V_{tg}$. The sharp drop in conductance around ±(0.2-0.4) meV corresponds to the superconducting transition of the Ta electrode, whose normal-state resistance is estimated to be $R_{N,Ta}$ = 107 Ω. Figure 2b shows expanded spectra of Figure 2a at low-bias voltages, where a zero-bias conductance *dip* is clearly evidenced and accompanied by two coherent-peak-like shoulders. The positions of the shoulders depend on $V_{tg}$; they move to higher bias voltage with increasing electron density in the InAs QW, which allows for larger critical current as shown in Figure 1c.

The observed d$I$/d$V$ spectral lineshape of transparent InAs−Ta junctions can be analyzed quantitatively using the standard BTK theory[22], after subtracting a common $R_{N,Ta}$ from the spectra taken at different $V_{tg}$ values. Explicitly, the interfacial conductance of the junction is given by

$$\frac{dI}{dV}(V) = \frac{Ne}{h}\int_{-\infty}^{\infty}\frac{df_0(E-eV)}{dV}[1 + A(E) - B(E)]dE, \quad (1)$$

where $N$ is the number of modes in InAs QW, $f_0(E)$ is the Fermi Dirac function, and $A(E)$ and $B(E)$ are probabilities of Andreev and ordinary reflection at the interface. The values of $A(E)$ and $B(E)$ at temperature $T$ depend on the superconducting gap $\Delta_T$ of Ta as well as the interfacial barrier strength $Z$ (dimensionless) with $Z$ = 0 for a perfect interface and $Z \rightarrow \infty$ in the tunneling regime. The dash lines in Figure 2b show best fits to the data using Eq. (1) and the corresponding fitting parameters are summarized in Table 1. A depairing parameter $\Gamma$ is also introduced as the complex component of the quasiparticle energy $E$, $E \rightarrow E + i\Gamma$, to take account of the inelastic scattering and achieve best fits[33]. We note that for three different $V_{tg}$, similar $Z$ values are obtained and $\Gamma \ll \Delta_{0.33K}$ (attesting the validity of our fitting). The obtained $Z$ values are comparable with the lowest reported value of 0.4 for Nb−InAs−Nb Josephson junctions[34]. In Figure 2c we plot the contour plots of d$I$/d$V$ for the transparent InAs−Ta junction with respect to dc bias voltage and temperature at $V_{tg} = 0$ V. Here, the color contrast (red) highlights the temperature evolution of the two shoulders in d$I$/d$V$ spectra, whose positions are correlated with the temperature dependence of the superconducting gap and following a Bardeen-Cooper-Schrieffer (BCS)-like behavior (solid line) with $\Delta_0 = 0.11$ meV and $T_c = 0.92$ K. The gap to critical temperature ratio of $\Delta_0/k_B T_c = 1.39$ is slightly smaller than the BCS value of 1.76. For comparison, we also plot the conventional BCS temperature dependence in Figure 2c with $\Delta_0 = 0.13$ meV $= 1.76\ k_B T_c$ and $T_c = 0.86$ K (dash line), where the $\Delta_0$ value is extracted from best BTK fit[35].

**Table 1.** BTK fitting parameters of the three dash lines shown in Figure 2b.

| $V_{tg}$ (V) | $R_{N,Ta}$ (Ω) | $\Delta_{0.33K}$ (meV) | $Z$ | $\Gamma$ (meV) |
|---|---|---|---|---|
| 3.6 | 107 | 0.16 | 0.45 | 0.045 |
| 0 | 107 | 0.13 | 0.42 | 0.015 |
| -4.8 | 107 | 0.11 | 0.43 | 0.002 |

Figure 3 presents the magnetic-field dependence of the d$I$/d$V$ spectra across a transparent and a relatively resistive InAs−Ta junction. For the transparent junction, the effect of magnetic field is simply closing the superconducting gap (Figure 3a), therefore the zero-bias conductance values remain nearly constant below the critical field (Figure 3c). The zero-bias conductance peak of the relatively resistive junction (Figure 3b), on the other hand, decreases substantially with increasing magnetic field (Figure 3d).

In conclusion, we show that the electronic transport across InAs−Ta junction depends on its interfacial (Schottky) barrier strength and the Andreev reflection process. When the interface is imperfect, the measured conductance spectra exhibit a broad peak at zero bias and the peak amplitude is sensitive to magnetic field, the bias voltage across the junction, and temperature. When a highly transparent InAs−Ta interface is achieved, however, a zero-bias conductance dip appears as well as two coherent-peak-like features forming at bias voltages corresponding to the superconducting gap of Ta. Our work thus demonstrates the need of achieving highly transparent interfaces in InAs/GaSb-based hybrid structures for studying the intriguing Andreev bound states in this recently discovered two-dimensional topological system.

## AUTHOR INFORMATION

**Corresponding Author**

*E-mail: zhigang.jiang@physics.gatech.edu; wpan@sandia.gov.

**Notes**

The authors declare no competing financial interest.

## ACKNOWLEDGMENT

We thank Rui-Rui Du for helpful discussions and Mark Overberg for wafer growth. This work is jointly supported by LDRD at Sandia and by the U.S. Department of Energy, Office of Science, Basic Energy Sciences, Materials Sciences and Engineering Division. Device fabrication was

# REFERENCES

(1) Fu, L.; Kane, C. L. *Physical Review Letters* **2008**, 100, 096407.

(2) Sau, J. D.; Lutchyn, R. M.; Tewari, S.; Das Sarma, S. *Physical Review Letters* **2010**, 104, 040502.

(3) Alicea, J. *Physical Review B* **2010**, 81, 125318.

(4) Stanescu, T. D.; Sau, J. D.; Lutchyn, R. M.; Das Sarma, S. *Physical Review B* **2010**, 81, 241310.

(5) Mourik, V.; Zuo, K.; Frolov, S. M.; Plissard, S. R.; Bakkers, E. P. A. M.; Kouwenhoven, L. P. *Science* **2012**, 336, 1003.

(6) Rokhinson, L. P.; Liu, X.; Furdyna, J. K. *Nat Phys* **2012**, 8, 795.

(7) Das, A.; Ronen, Y.; Most, Y.; Oreg, Y.; Heiblum, M.; Shtrikman, H. *Nat Phys* **2012**, 8, 887.

(8) Nayak, C.; Simon, S. H.; Stern, A.; Freedman, M.; Das Sarma, S. *Reviews of Modern Physics* **2008**, 80, 1083.

(9) Alicea, J. *Reports on Progress in Physics* **2012**, 75, 076501.

(10) Beenakker, C. W. J. *Annual Review of Condensed Matter Physics* **2013**, 4, 113.

(11) Volovik, G. E. *Jetp Lett.* **2009**, 90, 398.

(12) Linder, J.; Tanaka, Y.; Yokoyama, T.; Sudbø, A.; Nagaosa, N. *Physical Review Letters* **2010**, 104, 067001.

(13) Wimmer, M.; Akhmerov, A. R.; Medvedyeva, M. V.; Tworzydło, J.; Beenakker, C. W. J. *Physical Review Letters* **2010**, 105, 046803.

(14) Lutchyn, R. M.; Sau, J. D.; Das Sarma, S. *Physical Review Letters* **2010**, 105, 077001.

(15) Yang, M. J.; Wang, F.-C.; Yang, C. H.; Bennett, B. R.; Do, T. Q. *Applied Physics Letters* **1996**, 69, 85.

(16) Liu, C.; Hughes, T. L.; Qi, X.-L.; Wang, K.; Zhang, S.-C. *Physical Review Letters* **2008**, 100, 236601.

(17) Knez, I.; Du, R.-R.; Sullivan, G. *Physical Review Letters* **2011**, 107, 136603.

(18) Fu, L.; Kane, C. L. *Physical Review B* **2009**, 79, 161408.

(19) Knez, I.; Du, R.-R.; Sullivan, G. *Physical Review Letters* **2012**, 109, 186603.

(20) Nguyen, C.; Werking, J.; Kroemer, H.; Hu, E. L. *Applied Physics Letters* **1990**, 57, 87.

(21) Nguyen, C.; Kroemer, H.; Hu, E. L. *Physical Review Letters* **1992**, 69, 2847.

(22) Blonder, G. E.; Tinkham, M.; Klapwijk, T. M. *Physical Review B* **1982**, 25, 4515.

(23) Pan, W.; Klem, J. F.; Kim, J. K.; Thalakulam, M.; Cich, M. J.; Lyo, S. K. *Applied Physics Letters* **2013**, 102, 033504.

(24) Seo, Y.; Qin, Y.; Vicente, C. L.; Choi, K. S.; Yoon, J. *Physical Review Letters* **2006**, 97, 057005.

(25) Chien, C. J.; Chandrasekhar, V. *Physical Review B* **1999**, 60, 3655.

(26) Yang, F.; Ding, Y.; Qu, F.; Shen, J.; Chen, J.; Wei, Z.; Ji, Z.; Liu, G.; Fan, J.; Yang, C.; Xiang, T.; Lu, L. *Physical Review B* **2012**, 85, 104508.

(27) Knez, I.; Du, R. R.; Sullivan, G. *Physical Review B* **2010**, 81, 201301.

(28) Charpentier, C.; Fält, S.; Reichl, C.; Nichele, F.; Nath Pal, A.; Pietsch, P.; Ihn, T.; Ensslin, K.; Wegscheider, W. *Applied Physics Letters* **2013**, 103, 112102.

(29) Nichele, F.; Nath Pal, A.; Pietsch, P.; Ihn, T.; Ensslin, K.; Charpentier, C.; Wegscheider, W. *arXiv:1308.3128*.

(30) Sheet, G.; Mukhopadhyay, S.; Raychaudhuri, P. *Physical Review B* **2004**, 69, 134507.

(31) van Wees, B. J.; de Vries, P.; Magnée, P.; Klapwijk, T. M. *Physical Review Letters* **1992**, 69, 510.

(32) Andreev, A. F. *Soviet Physics Jetp-Ussr* **1964**, 19, 1228.

(33) Pleceník, A.; Grajcar, M.; Beňačka, Š.; Seidel, P.; Pfuch, A. *Physical Review B* **1994**, 49, 10016.

(34) Giazotto, F.; Grove-Rasmussen, K.; Fazio, R.; Beltram, F.; Linfield, E. H.; Ritchie, D. A. *Journal of Superconductivity* **2004**, 17, 317.

(35) We note that the extracted gap values from BTK fits appear larger than those indicated by the positions of the two coherent-peak-like shoulders in d$I$/d$V$ spectra.

**Figure Captions**

**Figure 1.** (a) Normalized $dV/dI$ spectra of two InAs−Ta junctions as a function of $I_{dc}$. One junction (black curve) has a zero-bias resistance ~38 Ω with an interfacial area of ~82 $\mu m^2$, while the other junction (red curve) is highly transparent, exhibiting a zero-bias resistance ~4.2 Ω with an interfacial area of ~140 $\mu m^2$. Inset: Schematic cross section of an InAs−Ta junction (not to scale). (b, c) Normalized $dV/dI$ spectra of the two junctions at different $V_{tg}$. Inset to (b): $dI/dV$ spectra of the relatively resistive junction near zero-bias voltage. Inset to (c): Scanning electron microscope image of a device and schematics of the contact configuration. All data were taken at $T = 0.33$ K.

**Figure 2.** (a) $dI/dV$ spectra of a transparent InAs−Ta junction in a relatively large dc bias voltage range and at different $V_{tg}$. (b) Expanded spectra of (a) near zero bias. Dash lines illustrate BTK fits to the data, as described in the main text. The data presented in (a) and (b) were taken at $T = 0.33$ K. (c) Contour plot of $dI/dV$ with respect to dc bias voltage and temperature at $V_{tg} = 0$ V. Color contrast highlights the evolution of the two coherent-peak-like shoulders in $dI/dV$ spectra, which follows a BCS-like gap dependence (solid line) with $\Delta_0 = 0.11$ meV $= 1.39\ k_BT_c$ and $T_c = 0.92$ K. Quantitative fitting using BTK theory leads to $\Delta_{0.33K} = 0.13$ meV with $Z = 0.42$ and $\Gamma = 0.015$ meV. For comparison, dash line shows the conventional BCS temperature dependence with $\Delta_0 = 0.13$ meV $= 1.76\ k_BT_c$ and $T_c = 0.86$ K.

**Figure 3.** Magnetic-field dependence of the $dI/dV$ spectra across (a) a transparent InAs−Ta junction and (b) a relatively resistive junction. (c) and (d) plot the magnetic-field dependence of the zero-bias conductance of the two junctions. All data were taken at $T = 0.33$ K and the magnetic field was applied perpendicular to the QWs.

**Figure 1**

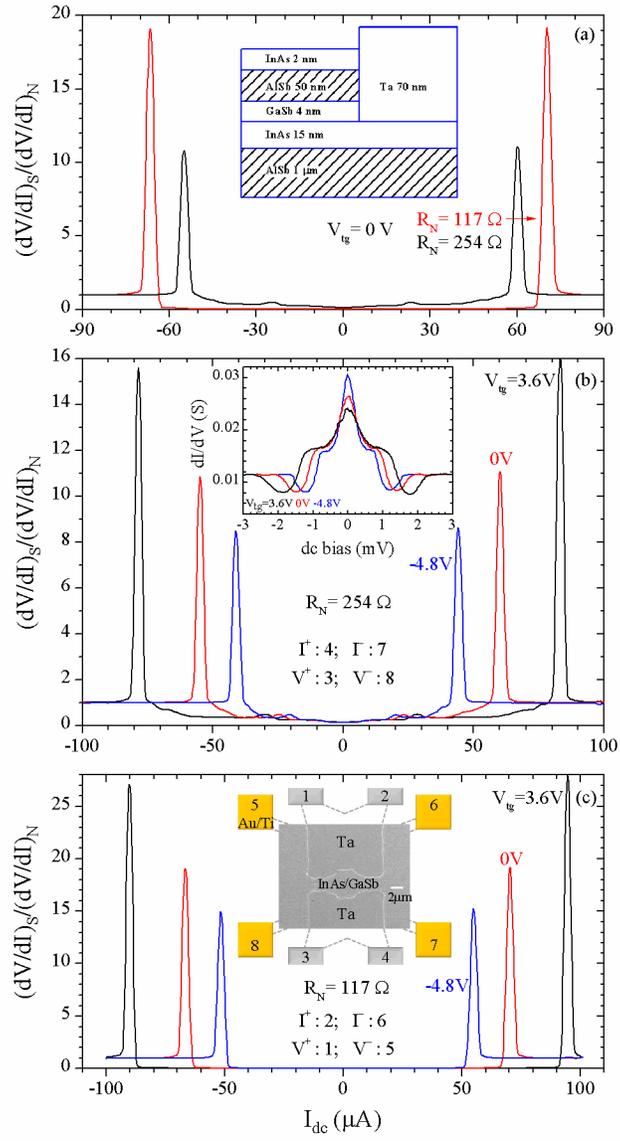

**Figure 2**

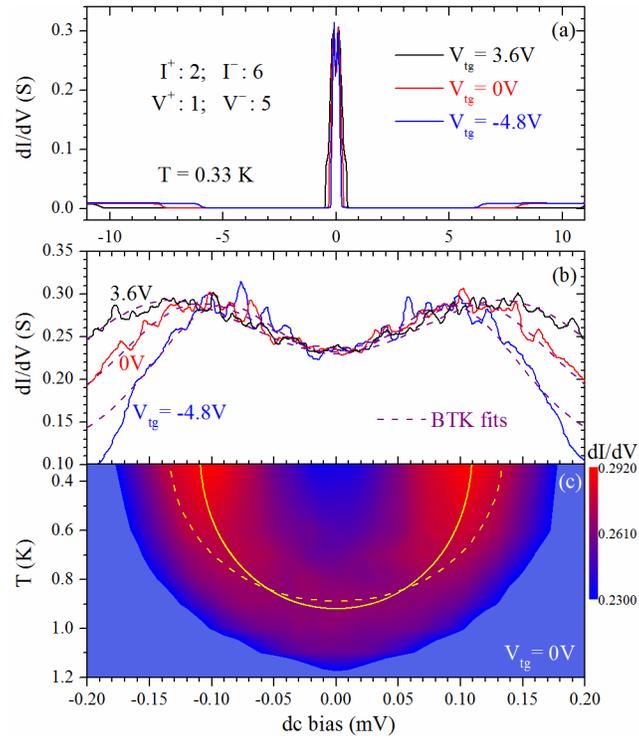

**Figure 3**

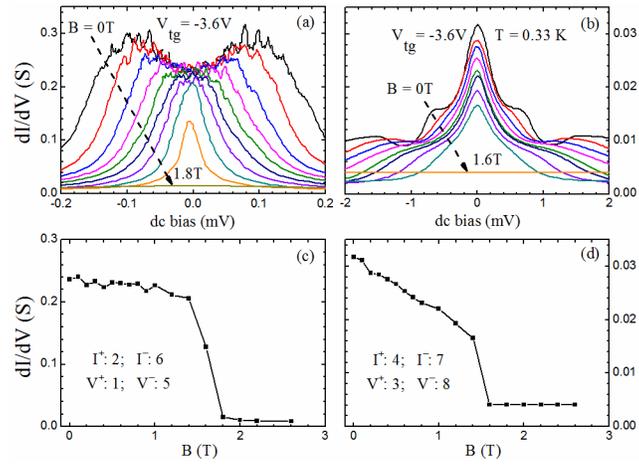